\documentclass[aps,twocolumn,prl,nofootinbib]{revtex4}
\usepackage{amsmath}
\usepackage{graphicx}
\usepackage{dcolumn}
\usepackage{bm}
\usepackage{amssymb}
\usepackage{latexsym}

\newcommand{\be}{\begin{equation}}
\newcommand{\ee}{\end{equation}}
\newcommand{\bq}{\begin{eqnarray}}
\newcommand{\eq}{\end{eqnarray}}

\begin{document}


\title{Comment on ``Turnaround in Cyclic Cosmology''}

\author{Xin Zhang}
\affiliation{College of Sciences, Northeastern University, Shenyang
110004, People's Republic of China} \affiliation{Kavli Institute for
Theoretical Physics China, Chinese Academy of Sciences, P.O.Box
2735, Beijing 100080, People's Republic of China}


\maketitle

Recently, it has been claimed that the entropy problem can be
resolved in a peculiar cyclic cosmology \cite{Baum:2006nz}. The
authors of the Letter proposed a spawn mechanism for helping
reconcile the infinitely cyclic cosmology with the second law of
thermodynamics. Here, we shall show that the spawn mechanism is
unlikely to be correct.

The Letter employed the phantom bounce model \cite{Brown:2004cs} as
a specific cyclic universe model. The Friedmann equation of this
model is given by
\begin{equation}
H^2={8\pi G\over 3}\rho\left(1-{\rho\over\rho_{\rm
c}}\right),\label{modiFeq}
\end{equation}
where $H=\dot{a}/a$ is the Hubble parameter, and $\rho_{\rm c}$ is
the critical energy density set by quantum gravity, which is the
maximal density of the universe. Such a modified Friedmann equation
with a phantom energy component leads to a cyclic universe scenario
in which the universe oscillates through a series of expansions and
contractions. In the usual universe, the phantom dark energy leads
to a ``big rip'' singularity; however, in this peculiar cyclic
universe, the big-rip singularity can be avoided because when $\rho$
reaches $\rho_{\rm c}$ the universe will turn around due to Eq.
(\ref{modiFeq}).

The authors of the Letter \cite{Baum:2006nz} argued that at the
turnaround point, our universe will be fragmented into a large
number of disconnected causal patches, each of which independently
contracts as a separate universe. The entropy of each causal patch
is essentially zero, i.e., $S={\cal O}(1)$ compared to the earlier
$S>10^{88}$. This dramatic decrease in entropy is called deflation
by the authors, which is claimed to provide a solution to the
entropy problem.

\begin{figure}[htbp]
\begin{center}
\includegraphics[scale=0.9]{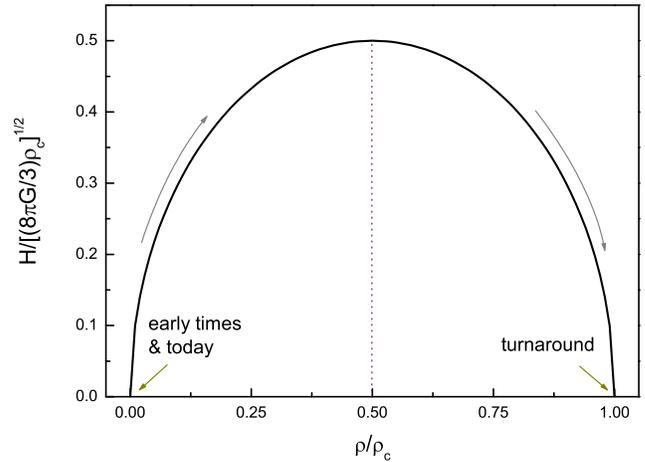}
\caption[]{\small Sketch map of the expanding phase of phantom
dominated universe in the cyclic cosmology.}\label{fig:turnaround}
\end{center}
\end{figure}

However, we shall show that this spawn mechanism (deflationary
hypothesis) is not realistic, i.e., our universe can not be
fragmented into independently causal patches at the turnaround.
According to Eq. (\ref{modiFeq}), though the phantom energy density
$\rho$ always increases with the expansion of the universe
\cite{note}, the Hubble parameter $H$ does not monotonously
increase, see Fig. \ref{fig:turnaround}. From Fig.
\ref{fig:turnaround}, it is clear that the Hubble parameter $H$
increases within the range $0<\rho<\rho_{\rm c}/2$ and decreases
within the range $\rho_{\rm c}/2<\rho<\rho_{\rm c}$; $H$ gets its
maximum at $\rho=\rho_{\rm c}/2$ (we call the corresponding time
$t_{\rm max}$). Turnaround occurs at a time $t_{\rm T}$, when
$H(t_{\rm T})=0$ (corresponding to $\rho=\rho_{\rm c}$).
Therefore, obviously, at the turnaround $t_{\rm T}$, the Hubble
length becomes infinity, $H^{-1}\rightarrow \infty$ (see also
\cite{Zhang:2007yu}). It is clear that though phantom energy makes
bound systems become unbound and the constituents causally
disconnect around $t_{\rm max}$, the many causally disconnected
patches reconnects together at the turnaround $t_{\rm T}$, if we
believe that the Hubble length plays an important role in
establishing the range of causal interaction or the size of a causal
region. So, we conclude that the universe would not be fragmented
into many disconnected causal patches at the turnaround.

It should be stressed that understanding the physics very near
turnaround is necessary to demonstrate the validity of the
fragmentation hypothesis. According to the analysis in this Comment,
the idea of Ref. \cite{Baum:2006nz} is albeit attractive but
unlikely to be correct. The entropy problem still exists in this
peculiar cyclic cosmology.

This work was supported by the China Postdoctoral Science
Foundation, the K. C. Wong Education Foundation, and the Natural
Science Foundation of China.


\begin{thebibliography}{99}

\bibitem{Baum:2006nz}
  L.~Baum and P.~H.~Frampton,
  Phys.\ Rev.\ Lett.\  {\bf 98}, 071301 (2007)
  [hep-th/0610213].


\bibitem{Brown:2004cs}
  M.~G.~Brown, K.~Freese and W.~H.~Kinney,
  astro-ph/0405353.

\bibitem{note}
We only consider the phantom dominated era in the expanding branch,
so the universe is in the high energy regime, $\rho \gg \rho_{\rm
today}$; we often say $\rho_{\rm today}\sim 0$ (hence $H_{\rm
today}\sim 0$).

\bibitem{Zhang:2007yu}
  X.~Zhang,
  arXiv:0708.1408 [gr-qc].





\end{thebibliography}
\end{document}